*Article*

# Changes in anthropogenic aerosols during the first wave of COVID-19 lockdowns in the context of long-term historical trends at 51 AERONET stations

Robert Blaga [1], Delia Calinoiu [2] and Gavrila Trif-Tordai [3,*]

[1] Faculty of Physics, West University of Timisoara, V. Parvan 4, 300223 Timisoara, Romania
[2] Department of Fundamental of Physics for Engineers, Politehnica University Timisoara, V. Parvan 2, Timisoara, Romania
[3] Faculty of Mechanical Engineering, Politehnica University Timisoara, M. Viteazu 1, Timisoara, Romania
* Correspondence: gavrila.trif-tordai@upt.ro

**Abstract:** A cvasi-consensus has steadily formed in the scientific literature on the fact that the prevention measures implemented by most countries to curb the 2020 COVID-19 pandemic have led to significant reductions in pollution levels around the world, especially in urban environments. Fewer studies have looked at the how these reductions at ground level translate into variations in the whole atmosphere. In this study, we examine the columnar values of aerosols at 51 mainland European stations of the Aerosol Robotic Network (AERONET). We show that when considered in the context of the long-term trend over the last decade, the columnar aerosol levels for 2020, at the regional level, do not appear exceptional. Both the yearly means and the number of episodes with extreme values for this period are within the one standard deviation of the long-term trends. We conclude that the spatially and temporally very localized reductions do not add up to statistically significant reductions in the global levels of aerosols. Furthermore, considering that pandemic lockdowns can be thought of as a simulation of a climate change mitigation scenario, we conclude that such lifestyle-based changes present a very low potential as a global climate change mitigation strategy.

**1. Introduction**

As the 2020 SARS-COV-2 pandemic swept across the world with frightening speed, governments attempted to manage its spread through partial or complete lockdowns. The restrictions were of varying levels – from voluntary social distancing implemented in Sweden to harsh stay-at-home lockdowns imposed in most countries – and were implemented and relaxed at different dates. Beyond 'flattening the curve' of hospitalizations, slowing human activities brought some welcome side effects. Global emission rates dropped, amounting to a predicted – notable, but not outstanding – a few percent decrease in $CO_2$ per 2020 [1,2]. A plethora of papers have been published reporting sharp declines in ground level anthropogenic pollutants such as $NO_2$ and particulate matter (PM), while $O_3$ generally increased [3-7]. In general, reductions were more visible in $NO_2$ levels, while PM reductions were more modest [8,9]. Furthermore, the reductions were shown to be large in cities where the average levels of pollution are also generally high, such as the rapidly urbanizing areas of India [10,11] or China [12], while in western countries they were less significant [13].

Some authors have also signaled some caveats to these findings. Ordóñez et al. [14] showed that care must be taken when interpreting results, because meteorological factors can outweigh emission cuts as drivers of changes in pollution levels. Wang et al. [15] noted that the pandemic restrictions are not enough to prevent severe air pollution events, especially in the presence of unfavorable weather. Most importantly, when looking at long-term trends, some authors have found that reductions for 2020 do not appear consistently across studied locations, and, where they do, they can be modest and transient [16-19].

As several authors have also noted [20], due to the severity and breadth of lockdown measures, our global reaction to the SARS-CoV-2 pandemic acts as a kind of artificial experiment or simulation of what a moderate climate-mitigation scenario – that focuses heavily on lifestyle changes – would look like. In this vein, the changes in greenhouse gas emissions in the context of the COVID-19 pandemic and the varied, but universal, reaction of countries give an assessment of the impact such mitigation measures could achieve.

To assess global emissions, we must look at columnar values of various atmospheric species. It is an interesting question whether the ground level reductions also translate into reductions in the bulk, once diffusion times are taken into consideration. Several authors have approached this issue with differing results. Ranjan et al. [21] have found significant reductions in aerosol optical depth (AOD) in various Indian locations. Venter et al. [22] have found that in general the sign of changes in the bulk for $NO_2$, $O_3$ and aerosols is the same as at the ground level, although their magnitude is more modest. Song et al. [23] found that clean air regulations and COVID-19 restrictions are competing factors that influence reductions in tropospheric $NO_2$ in several cities in East China.

Decreases in pollutant levels at the ground clearly have significant potential environmental and health benefits. Additionally, if reductions are also visible in the bulk, the slowing of society could have a positive impact on photovoltaic yields, by clearing urban air [24-26]. It is known that, in addition to clouds, aerosols represent the most potent atmospheric attenuator of the incoming solar irradiance. For the purpose of this study aerosols are used interchangeably with particulate matter (PM), both referring to suspended solid or liquid particles in the atmosphere. Calinoiu et al. [27] have shown that even with moderate aerosol loading, the losses in the collectable solar energy due to aerosol particles can exceed 20%.

The atmospheric extinction of solar irradiance is controlled by the columnar values of the relevant atmospheric species, quantified through the optical depth (OD) parameter. For a given atmospheric species, the OD is related to the corresponding atmospheric transmittance as follows:

$$\tau_i = exp(-m_i \cdot OD_i) \simeq 1 - m_i \cdot OD_i, \qquad (1)$$

where $i$ designates the atmospheric species and $m_i$ represents the optical mass of species $i$. The second equality holds for small values of the parameter $x_i = m_i \cdot OD_i$. Thus, for small values, the parameter $x_i$ represents the fraction of the incident solar beam that is lost (i.e., scattered or absorbed) due to species $i$. The optical mass is unity for the sun positioned at the zenith, and grows to above 10 close to sunrise/sunset. If $m$ is modeled identically for all species, the total optical depth is simply the sum of the individual ODs. In the absence of clouds, aerosols represent the most impactful species in regard to atmospheric extinction in terms of both absolute magnitude and variability. Other impactful species are water vapor (w), ozone ($O_3$), and nitrogen-dioxide ($NO_2$).

In this study, we focus on the atmospheric level of aerosols at a number of mainland European stations from the Aerosol Robotic Network (AERONET). AERONET is currently considered the most accurate source of ground measurements - particularly for AOD – widely used as a benchmark for validating satellite data products or reanalysis [28]). Measurements were performed using Cimel Electronique Sun-sky radiometers. Columnar values and associated optical depths for various atmospheric species have been investigated in the last decade, with a focus on aerosols. A number of 51 stations were selected, which had archived values in 2020. From these stations, data from the period 2010-2023 is used here as a working dataset. The stations – with their AERONET name – are listed and represented in Figure 1.

Studies that find reductions in pollution levels almost exclusively refer to concentrations or mass density at ground level. Anthropogenic pollutants have circulation times from days to a few weeks [29]. Thus, it is natural for there to be a delay in the time that reductions become visible in measurements in the bulk (columnar atmospheric values) as compared to values at ground level. However, even taking into consideration this delay, we find that specifically for aerosols the variations in columnar values for the pandemic period are within the limits of inter-annual variability, and thus cannot be statistically strongly concluded to be driven by the COVID-19 lockdowns.

## 2. Dataset and Methods

Our data set comprises ground measurements from 51 stations from the Aerosol Robotic Network (AERONET). The stations are represented in Figure 1. At AERONET stations, data is recorded with 15-min resolution. The version 3, level 2 datasets are used here, which contain data points recorded under only clear sky conditions. In this study, data for the period 2010-2023 is used for each station. Only the months of March, April, May and June, covering the first wave of the SARS-CoV-2 pandemic, were retained for each year. Henceforth, if "yearly" values are mentioned or represented, they refer to values over this period. Furthermore, although the pandemic had several subsequent waves, the most stringent lockdown measures were implemented in each country during the first months. Thus, throughout this study we refer to these four months as the "pandemic period"

The stations were grouped into four regions: Southern Europe (SEU: Italy, Portugal, and Spain), Central Europe (CEU: Austria, Germany, and Switzerland), Eastern Europe and the Balkans (EEB: Belarus, Estonia, Greece, Poland, Romania, Russia, Slovakia, and Ukraine), and Western Europe (WEU: France and Belgium). This partitioning has nothing to do with historical or political geographies, it is a simple loose grouping of countries that are geographically close and have had a somewhat similar approach to the pandemic. The latter argument is the reason why, for example, Greece was included in the EEB group instead of the SEU. It was also considered that the regions have a balanced number of stations and data points.

Data lines with missing or obviously unphysical values for at least one of the relevant parameters have been removed from the datasets. In order to eliminate days with predominantly desert dust – the main type of non-anthropogenic tropospheric aerosol – the condition that the Ångström Exponent be higher than 1.3 was imposed [30,31]. The Ångström Exponent is inversely related with the particle diameter, and hence, this condition eliminates all coarse mode particles. A more detailed description of the Ångström Exponent is given in Section 3.1.

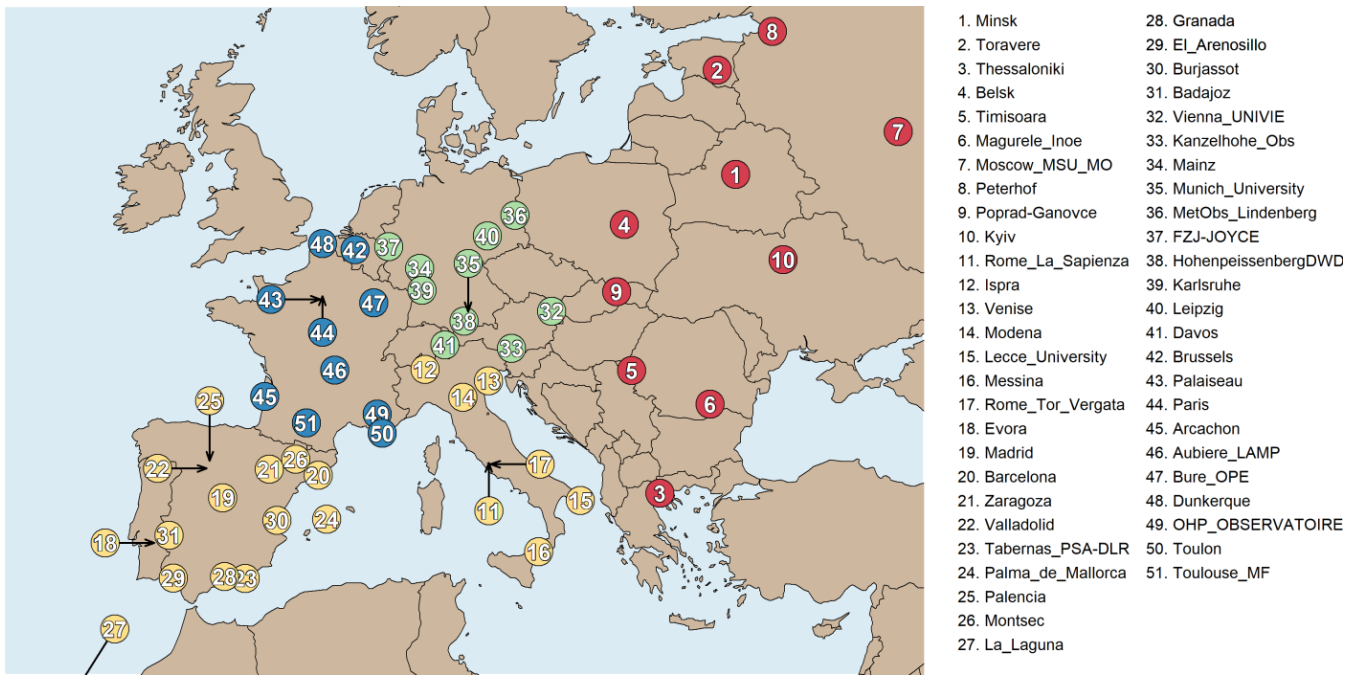

**Figure 1.** Stations included in this study, listed according to their names from AERONET.

*2.1. Inter-scenario comparability*

When comparing aerosol data recorded at different locations or at different times, one must control for external factors. In the case of ground-level in situ measurements, differences in weather and humidity need to be accounted for, as these strongly impact aerosol properties. In the case of AERONET data, the retained measurements are all under clear-sky conditions, which automatically guarantees a certain degree of similarity between scenarios. We use an additional proxy to check statistical similarity in weather conditions: the average precipitable water column content (PW). The PW represents the quantity of water that is present in the atmosphere in vapor form. Sulfates and other inorganic aerosols absorb water above a relative humidity threshold, changing their physical and optical properties. Thus, in terms of aerosol growth and dynamics, PW is a very impactful quantity. Generally, PW levels, perhaps unintuitively, have a strong positive correlation with temperature and are thus also a marker of heatwaves or cold spells. PW levels are assessed at the regional level, as defined in the previous section.

Figure 2 shows the PW anomalies for the four regions during the period studied. Measured values for 2020 are compared to the standard deviation (σ) over the period 2010-2019 at each station. We observe that 2020 is not an exceptional year in regard to PW, none of the recording differing by more than one standard deviation compared to the long-term trends. There are also no notable large-scale features in the left panel of Figure 2 besides the low-PW episode during the month of May at EEB locations. Two extreme examples are shown, for comparison, in the right panel of Figure 2. Namely, the low temperatures and PW levels associated with prolonged winter in 2013, and the high temperatures and PW associated with the 2018 heatwave.

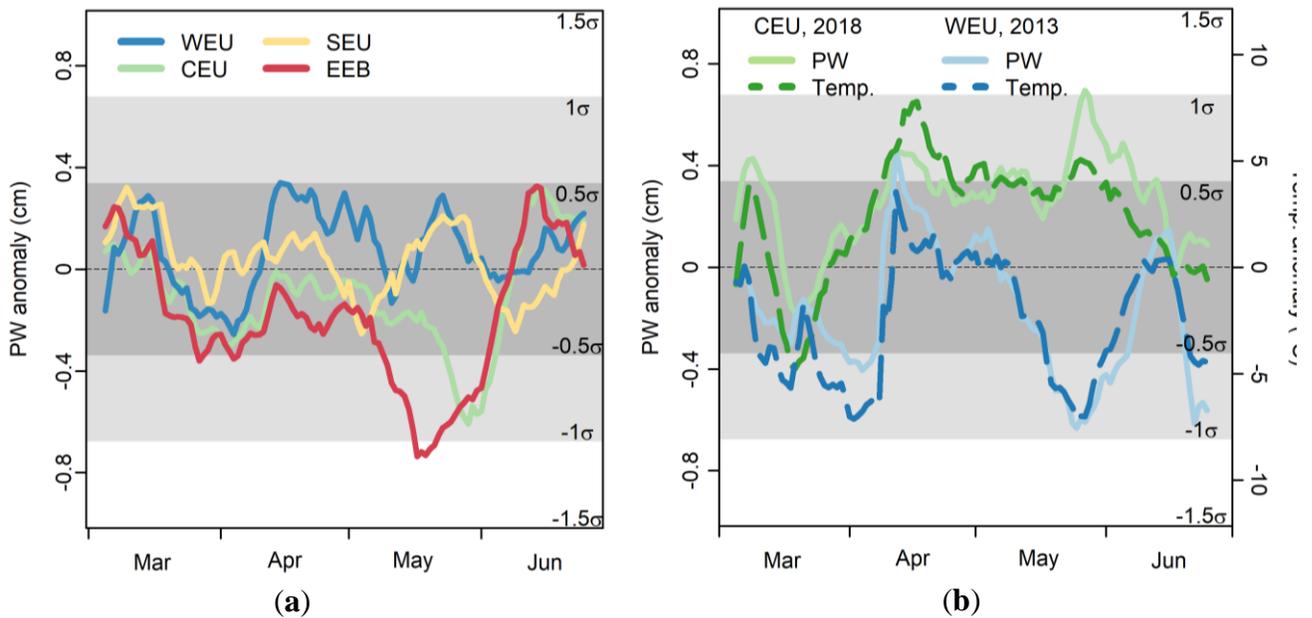

(a)  (b)

**Figure 2.** (**a**) 10-day moving average of the daily PW anomaly for 2020 over the considered period. The anomalies are calculated in relation to the averages for the given day over the period 2010-2019. The standard deviation ranges for the entire dataset are also represented in the background; (**b**) PW and temperature anomalies for two particular cases from 2013 and 2018.

*2.2. Restrictions calendar*

Different countries responded to the health crisis with different combinations of restrictions. Most European countries imposed stay-at-home lockdowns (besides the notable exceptions of Sweden, Belarus,

Iceland and Latvia). In addition to this and other common measures, such as closing borders and internal travel restrictions, there is no universal set of regulations imposed. Table 1, compiled from a variety of sources, provides a table of restriction levels [32,33]. It must be noted that the categorization of measures listed in Table 1 has a high degree of subjectivity, given the non-uniformity of the set of measures imposed per country or even within one country. Still, it can be seen that for most countries, the stage of harshest restrictions spans mid-March to mid-May. In this phase, restrictions often included closure of schools, stay-at-home lockdowns and encouragement or imposition of work from home where possible. During the height of the lockdowns, air traffic also was drastically reduced.

**Table 1**. Level and period of restrictions for each country included in this study. Shades mark the level (1-4) of restrictions (light to dark). Restriction levels are based loosely on the description from [34].

| | Feb | Mar | | | | Apr | | | | May | | | | Jun | | | | Jul |
|---|---|---|---|---|---|---|---|---|---|---|---|---|---|---|---|---|---|---|
| | Q4 | Q1 | Q2 | Q3 | Q4 | Q1 | Q2 | Q3 | Q4 | Q1 | Q2 | Q3 | Q4 | Q1 | Q2 | Q3 | Q4 | Q1 |
| Belgium | | | | | | | | | | | | | | | | | | |
| France | | | | | | | | | | | | | | | | | | |
| Austria | | | | | | | | | | | | | | | | | | |
| Germany | | | | | | | | | | | | | | | | | | |
| Switzerland | | | | | | | | | | | | | | | | | | |
| Italy | | | | | | | | | | | | | | | | | | |
| Portugal | | | | | | | | | | | | | | | | | | |
| Spain | | | | | | | | | | | | | | | | | | |
| Belarus | | | | | | | | | | | | | | | | | | |
| Estonia | | | | | | | | | | | | | | | | | | |
| Greece | | | | | | | | | | | | | | | | | | |
| Poland | | | | | | | | | | | | | | | | | | |
| Romania | | | | | | | | | | | | | | | | | | |
| Russia | | | | | | | | | | | | | | | | | | |
| Slovakia | | | | | | | | | | | | | | | | | | |
| Ukraine | | | | | | | | | | | | | | | | | | |
8

## 3. Analysis

The presence of nitrogen oxides (NO and $NO_2$), and of certain volatile organic compounds (VOCs) in high mixing ratios is a clear marker of human activity. In urban environments, values many times above the natural continental background are usually registered. In addition to their direct impact on human health when breathed in, they represent precursor species for the formation of $O_3$ and urban-industrial particulate matter, with its most well-known manifestation – smog. As detailed in Introduction, many authors have found significant reductions in both $NO_x$ levels and PM at the ground during the pandemic months. In this section, we investigate whether these ground-level reductions in pollutants due to the COVID-19 restrictions in the period March-June 2020, lead to decreases in the columnar values of aerosols to a statistically significant degree during this period.

### 3.1. Aerosols and urban pollution

One of the main drivers of negative health outcomes and the main atmospheric solar attenuator is represented by suspended particulate matter (PM) in the atmosphere. One measure of the abundance of such atmospheric aerosols is the Ångström turbidity coefficient ($\beta$). $\beta$ is related to the aerosol optical depth through the Ångström relation, which – in its simplest form – is written as:

$$AOD(\lambda) = \beta \lambda^{-\alpha} \qquad (2)$$

where $\lambda$ represents wavelength expressed in µm, and $\alpha$ is the Ångström exponent. The AOD is written in this form in order to obtain information about aerosol composition and concentration. While $\alpha$ and $\beta$ are not directly observable quantities, it can be shown that $\alpha$ is related to the inverse of the aerosol dynamic diameter, while information about the particle number is captured in $\beta$. $\alpha$ can be obtained by writing Equation (2) for two wavelengths $\lambda_1$, $\lambda_2$ and then diving to eliminate $\beta$. The resulting relation can be inverted to obtain:

$$\alpha = -\ln\frac{\lambda_2}{\lambda_1} \ln\frac{AOD(\lambda_1)}{AOD(\lambda_2)} \qquad (3)$$

As is practice in the literature, the values 0.44µm and 0.87µm are used in Equation (3). $\beta$ is then obtained from the AOD at 0.5µm.

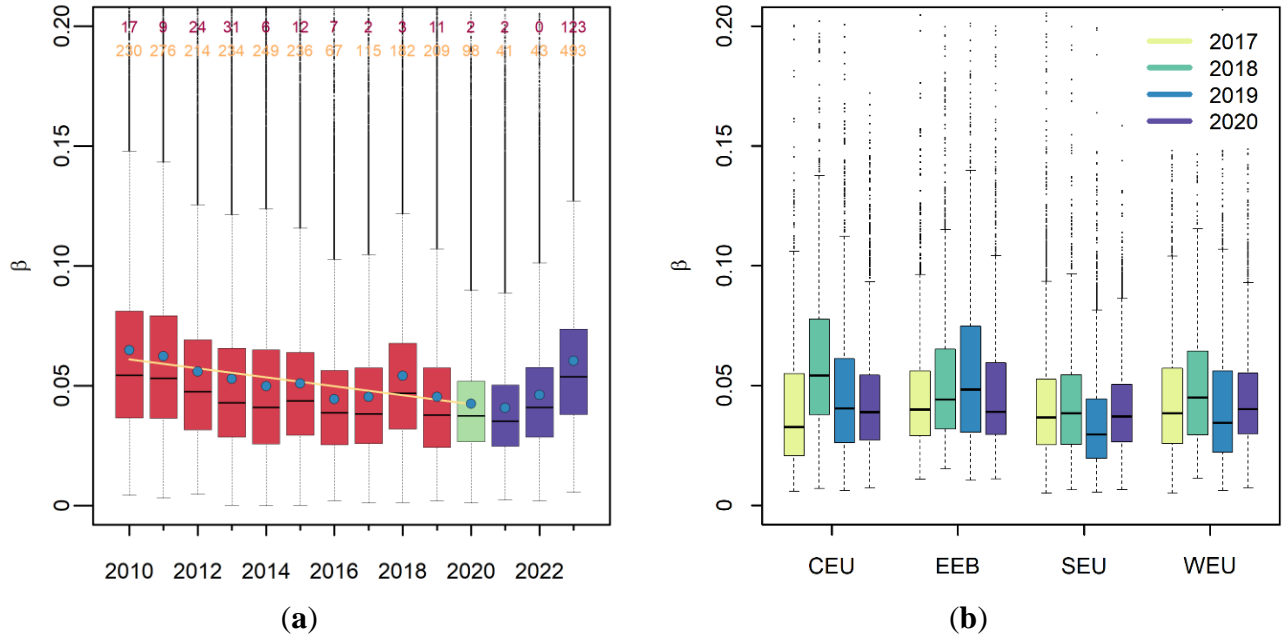

**Figure 3.** (**a**) Boxplot of the Ångström turbidity coefficient per year for our dataset. The midlines in the boxes represent median values, while the purple dots represent averages. The yellow line is the linear trend for 2010-2019. Numbers on the top part represent the number of episodes with moderate to high (orange, $\beta > 0.2$) and extreme (dark red, $\beta > 0.5$) pollution; (**b**) Yearly boxplots of the Ångström turbidity during the period 2017-2020 for each major region.

For point sources at the ground level, the horizontal distribution of aerosols falls-off exponentially with air density, on the length scale determined by the planetary boundary layer height. For these reasons, the major contribution to the column-integrated aerosol properties comes from aerosols close to the ground. Thus, the Ångström turbidity parameter $\beta$ can be seen as indicative of aerosol loading levels close to the ground. Furthermore, because the information on the concentration of particle numbers is encapsulated in $\beta$, it's study can be relevant to health studies.

Yearly boxplots of $β$ across the entire data set are shown in Figure 3. Qualitatively, a general decreasing trend is visible for the decade 2010-2019, with low points during 2020-2021, and a sharp increasing trend during 2022-2023. A decreasing trend of -0.00187/yr ± 0.00405/yr for $β$ is found for the period 2010-2019. The predicted value for the 2020 mean $β$ based on the linear trend is 0.04228, while the measured value is 0.04249. The difference is well within the standard deviation of the trend residuals.

The situation is similar when we look at the annual number of moderate or extreme pollution. In line with the categorization from Blaga et al. [35], values of β below 0.08 are considered indicative of clean air (continental background aerosol levels), moderate aerosol loading is considered at values around 0.1-0.2, while a value above 0.5 means an extreme pollution event. The number of moderate and severe pollution episodes per year is shown in Figure 3a. A trend of -11.2/yr ± 54.4/yr for moderate episodes is found, for which the 2020 value of 98 is within the standard deviation of the trend. For extreme episodes, the trend is -1.6/yr ± 7.9/yr, for which the 2020 value of 2 is, again, within the trend.

Therefore, we conclude that there is no statistically significant variation for the 2020 values. Both the average values and the frequency of high pollution episodes for 2020 are consistent with the multiannual decreasing trend. In the right panel, we see the breakdown of β per region, for the years 2017-2020. It is interesting to note how different the interpretation of the 2020 values is if we look only at the last 2-3 years. For CEU locations, 2020 seems to fall within the decreasing trend, for example, while at EEB locations 2020 seems to break with the increasing trend. However, in the bigger picture, the values are all consistent with the long-term decreasing trend.

*3.2. Aerosols and photovoltaic yields*

Fundamentally, the yield from a photovoltaic cell is controlled by the amount of solar irradiance that is incoming on its surface, which in turn is determined by the attenuation processes that occur in the atmosphere, modelled as a transmittance for each atmospheric species through Equation (1).

In the case of Si photovoltaic cells, the spectral response is broad and thus a broadband AOD is more useful for gauging photovoltaic losses instead of the spectral AOD. The broadband AOD is introduced here through spectral averaging-based on the AERONET data and Equation (2) - as follows:

$$\overline{AOD} = \frac{1}{\int G_0(\lambda)d\lambda} \int G_0(\lambda) AOD(\lambda) d\lambda \qquad (4)$$

where $G_0$ is the solar irradiance in the absence of the atmosphere, used here as the AM0 spectrum. The Ångström parameters $\alpha$ and $\beta$ are determined using Equation (2) based on the measured AERONET AOD at 0.5µm. The reader is reminded that for low-to-moderate aerosol loading, the AOD is approximately equal to the fraction of the incident solar beam lost through scattering and absorption along the direction of the beam, as explained in Equation (1).

Yearly broadband AOD is shown in the form of bar graphs in Figure 4, for the four regions. A one standard deviation band is also shown, which helps with the visual identification of outlier years. The situation is different for different regions. The WEU locations show the fastest reduction in aerosol levels with a slope of -0.0071/yr, however the trend is distorted partly by very high AOD values during 2010-2012. The SEU and EEB locations experience a milder decreasing trend, with slope equal to -0.0036/yr and -0.0027/yr. The CEU region is peculiar, having a very small slope of -0.0016/yr, but a high volatility, the standard deviation of residuals being equal to 0.0154 (highest of all four regions). However, some universal results were found across all four regions:

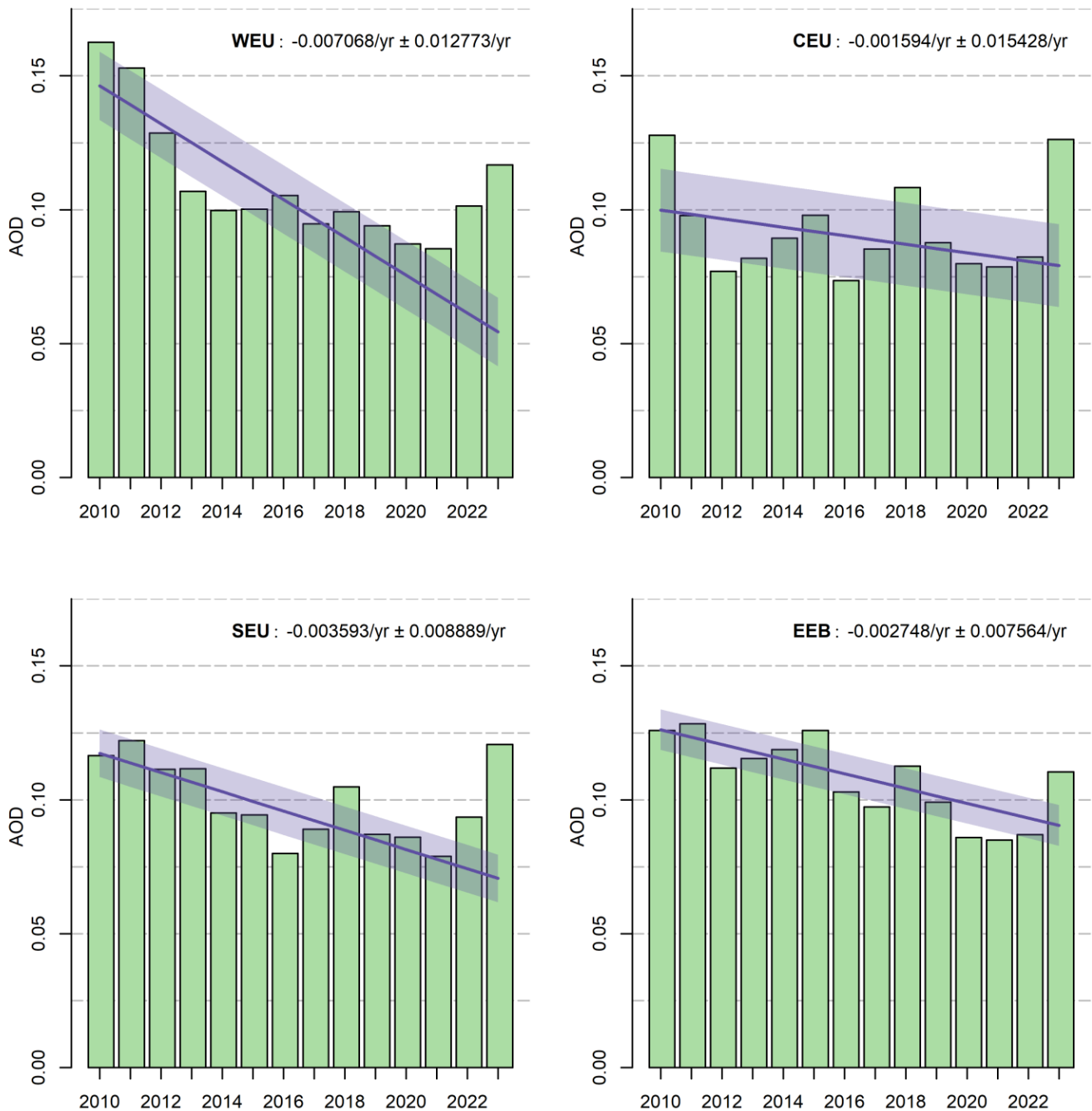

**Figure 4**. Broadband Aerosol Optical Depth per year on the four studied data regions. Purple lines represent the linear trend, while the purple band spans one standard deviation of the residuals above and below the trend. The slope and standard deviation are also listed numerically.

i) the years 2010-2011 were characterized by relatively high aerosol loading;

ii) a general decreasing trend present during the interval 2010-2019, influenced by heightened air quality policies across EU countries;

iii) the year 2023 is an outlier with very high aerosol levels, with more than one standard deviation above the multiyear linear trend;

iv) the year 2020 is within one standard deviation of the multiyear trend, meaning it is not a statistical outlier.

The last point parallels our hypothesis that the pandemic year 2020, when considered in terms of columnar aerosol loading, is not an outlier. To further investigate this hypothesis, the standardized AOD anomaly for 2020 is calculated. For each station, a linear fit is performed on the period 2010-2019. The anomaly is then calculated as:

$$\delta\overline{AOD}_{2020} = \left(\overline{AOD}_{2020} - \widehat{AOD}_{2020}\right)/\sigma_{2010-2019} \tag{3}$$

where $\overline{AOD}_{2020}$ represent the 2020 broadband AOD, while $\widehat{AOD}_{2020}$ is the same optical depth estimated with the linear fit. $\sigma$ represents the standard deviation of the residuals of the linear trend over the fit period 2010-2019. The 2020 AOD anomaly for our database is shown in Figure 5. We observe that for most stations the 2020 AOD falls within one standard deviation of the decreasing linear trend. Taking into account the working hypothesis in this study, 12/51 stations fall more than one standard deviation, and 5/51 stations fall more than two standard deviations below the trend. The stations with the most extreme anomalies for the pandemic year 2020 are, in order: Modena, Italy; Toulon, France; Poprad-Ganovce, Slovakia; Dunkerque, France; Belsk, Poland. Locations do not cluster geographically nor regionally. The two stations with the largest relative reductions for 2020 are ones that generally experience the highest average AOD (~0.15). However, this does not turn out to be a general rule, as stations with a lower AOD (~0.1) are also among the handful of negative outliers.

In conclusion, the 2020 AOD levels are not statistical outliers to a degree that the current study over 51 European AERONET stations can show. We interpret this result in the following way: it seems that the ground-level reductions in air pollutants resulting from the COVID-19 pandemic-induced lockdowns were not persistent enough to translate into reductions in the columnar aerosol levels. Furthermore, based on Figure 4 and 5, we can also conclude that there have been no improvements in photovoltaic yields across mainland Europe, that can be statistically strongly linked to the COVID-19 restrictions.

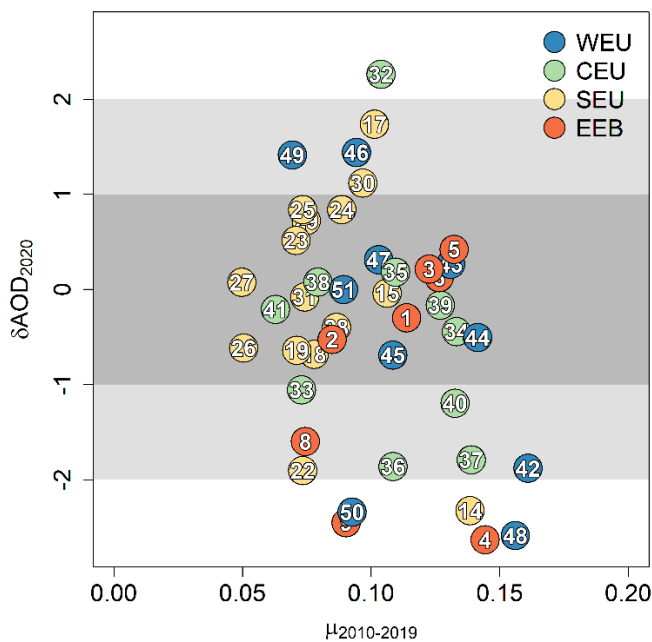

**Figure 5**. Standardized 2020 AOD anomaly [Equation (5)] versus mean AOD for the period 2010-2019, for each station from our database. The station regions are color-coded. The stations are numbered according to Figure 1.

**4. Summary and Conclusion**

In this study, a database of ground-measurements from various mainland European locations of the AERONET network was used to analyze the columnar levels of aerosols during the period marked by the first wave of the COVID-19 pandemic.

No significant reductions in columnar values were found, despite the sharp reductions in ground-level emissions reported in urban areas by many authors. In fact, variations of Ångström turbidity and broadband aerosol optical depth are within the margins of the multi-year trends and, thus, no statistically significant improvement of any kind concerning emissions is observable, at least at the studied locations, besides the very localized – in both space and time – reductions in values at ground level.

These results, together with the relatively small reduction in $CO_2$ emissions, in fact show the small contribution mundane anthropogenic activities have to overall greenhouse gas emissions. Even regions where non-essential economic activity was halted did not see dramatic decreases. We can only speculate on the cause, but it seems probable that globally the main sources – as is the case with $CO_2$ [36] – are large individual economic actors; actors that have not stopped or reduced their activities during the pandemic.

Unfortunately, the COVID-19 restrictions also did not lead to increased yields from photovoltaic cells. As we have seen, particulate matter levels in the troposphere – the largest attenuator, except clouds – have not been significantly affected. Considering also the drop in sales and installed capacity during this period [37] – from all sources, not just solar – we can conclude that the pandemic period was exclusively harmful to the renewable energy sector.

It can be argued that the COVID-19 restrictions are a simulation of a climate change mitigation scenario centered on lifestyle changes and moderate economic restrictions. Viewed through this lens, the results from our analysis paint a very pessimistic picture about the possibilities for avoiding the worst climate change scenarios, in the absence of a large-scale reorganization of our socio-economic world-system.

**Author Contributions:** Conceptualization, R.B. and C.D.; methodology, R.B.; validation, R.B. and C.D.; formal analysis, C.D. and G.T.; investigation, R.B., C.D. and G.T.; resources, G.T.; data curation, R.B.

and G.T.; writing-original draft preparation, R.B.; writing-review and editing, R.B., C.D. and G.T. All authors have read and agreed to the published version of the manuscript.

**Data Availability Statement:** The original contributions presented in the study are included in the article, further inquiries can be directed to the corresponding author.

**Acknowledgments:** The authors express their gratitude to the personnel who maintain the 51 AERONET stations considered in this study.

**Conflicts of Interest:** The authors declare no conflicts of interest.